\documentclass[twocolumn,amsmath,amssymb,prl]{revtex4}

\usepackage{graphicx}
\usepackage{dcolumn}
\usepackage{bm}

\newcommand{\rme}{\mathrm{e}}

\begin{document}

\preprint{APS/123-QED}

\title{Comment on ``Nonlocal Coherent Perfect Absorption''}

\author{Pablo L. Saldanha}\email{saldanha@fisica.ufmg.br}
\affiliation{Departamento de F\'isica, Universidade Federal de Minas Gerais, Belo Horizonte, MG 31270-901, Brazil}


\date{\today}


\maketitle

In Ref. \cite{jeffers19}, Jeffers discuss interesting behaviors of photons propagating through an arrange of lossy beam splitters that compose an interferometer. The system permits the control of which input modes are absorbed with certainty in the interferometer, and which input modes are not absorbed with maximum probability. However, the correct interpretation of the phenomenon is not as stated in this reference. As discussed here, there is no `nonlocal coherent perfect absorption' by the lossy beam splitters $a$ and $b$ of Fig. 1(a) of Ref. \cite{jeffers19}, as stated in the title of the paper, but a local coherent absorption by the other two lossy beam splitters in this figure. 

To introduce our notation, that evolves the system state instead of the operators, first consider a single photon interacting with a lossy beam splitter as in Fig. 1(b) of Ref. \cite{jeffers19}. We have $|1\rangle\equiv\hat{a}^\dag_\mathrm{1in}|\mathrm{vac}\rangle$ and $|2\rangle\equiv\hat{a}^\dag_\mathrm{2in}|\mathrm{vac}\rangle$, where $|\mathrm{vac}\rangle$ represents the vacuum state for all relevant modes and the notation of Eq. (1) of Ref. \cite{jeffers19} is used. The beam splitter output modes are also denoted $|1\rangle$ and $|2\rangle$. If the beam splitter absorbs a photon, it evolves to state $|f_1\rangle$ or $|f_2\rangle$ depending on the input port of the photon. We have
\begin{equation}
	\langle f_1|f_2\rangle \equiv \langle \mathrm{vac}|\hat{f}_{1}\,\hat{f}_{2}^\dag|\mathrm{vac}\rangle=\langle \mathrm{vac}|\Big([\hat{f}_{1},\hat{f}_{2}^\dag]+\hat{f}_{2}^\dag\hat{f}_{1}\Big)|\mathrm{vac}\rangle.             
\end{equation}
According to the text below Eq. (1) in Ref. \cite{jeffers19}, we have $[\hat{f}_{1},\hat{f}_{2}^\dag]=-2tr/{l^2}$ for  real coefficients $r$, $t$, and $l$, such that
\begin{equation}\label{scalar}
	\langle f_1|f_2\rangle=-\frac{2tr}{l^2}.
\end{equation}
With the input state $(|1\rangle+\rme^{i\phi}|2\rangle)/{\sqrt{2}}$ at the beam splitter, the output state is
\begin{equation}
	\frac{t+r\rme^{i\phi}}{\sqrt{2}}|1\rangle+\frac{r+t\rme^{i\phi}}{\sqrt{2}}|2\rangle+\Bigg[\frac{l}{\sqrt{2}}|f_1\rangle+\frac{l\rme^{i\phi}}{\sqrt{2}}|f_2\rangle\Bigg].
\end{equation}
Using Eq. (\ref{scalar}), the probabilities $P_1$, $P_2$, and $A$ of photon detection in modes 1 and 2 and absorption by the beam splitter, respectively, are
\begin{equation}
	P_1=P_2=\frac{t^2+r^2}{2}+rt\cos(\phi)\,, \; A=l^2-2rt\cos(\phi),
\end{equation}
which sum to 1 and may present coherent perfect absorption and transparency when $t=\pm r=1/2$ and $l=1/\sqrt{2}$ \cite{jeffers00}.

Now let us discuss the scheme of Fig. 1(a) of Ref. \cite{jeffers19}, denoting the beam splitter that couples light to detectors D$_3$ and D$_2$ (D$_1$ and D$_4$) as $c$ ($d$), with coefficients $t_c$, $r_c$ and $l_c$ ($t_d$, $r_d$ and $l_d$). As in Ref. \cite{jeffers19}, the input state has one photon in a superposition of being in modes 1 and 3: $(|1\rangle+\rme^{i\phi}|3\rangle)/\sqrt{2}$. The system state just before beam splitters $c$ and $d$ is
\begin{eqnarray}\nonumber
	|\Psi_1\rangle&=&\frac{1}{\sqrt{2}}\Big[t_a|1\rangle+r_a|2\rangle +l_a |f_{a1}\rangle \Big]+\\
	                &+&\frac{\rme^{i\phi}}{\sqrt{2}}\Big[t_b\rme^{i\theta_3}|3\rangle+r_b\rme^{i\theta_4}|4\rangle +l_b|f_{b2}\rangle \Big].
\end{eqnarray}
The system state after beam splitters $c$ and $d$ is 
\begin{eqnarray}\label{psi2}\nonumber
	|\Psi_2\rangle&=&\frac{t_at_d+r_br_d\rme^{i(\phi+\theta_4)}}{\sqrt{2}}|1\rangle+\frac{r_at_c+t_br_c\rme^{i(\phi+\theta_3)}}{\sqrt{2}}|2\rangle+\\\nonumber
	              &+&\frac{r_ar_c+t_bt_c\rme^{i(\phi+\theta_3)}}{\sqrt{2}}|3\rangle+\frac{t_ar_d+r_bt_d\rme^{i(\phi+\theta_4)}}{\sqrt{2}}|4\rangle+\\\nonumber 
								&+&	\frac{l_a}{\sqrt{2}} |f_{a1}\rangle+\Bigg[\frac{r_al_c}{\sqrt{2}}|f_{c1}\rangle+	\frac{t_bl_c\rme^{i(\phi+\theta_3)}}{\sqrt{2}}|f_{c2}\rangle\Bigg]+              \\
								&+&	 \frac{l_b\rme^{i\phi}}{\sqrt{2}} |f_{b2}\rangle +\Bigg[\frac{t_al_d}{\sqrt{2}}|f_{d2}\rangle+	\frac{r_bl_d\rme^{i(\phi+\theta_4)}}{\sqrt{2}}|f_{d1}\rangle\Bigg].
\end{eqnarray}

Considering $t_i=-r_i=1/2$ and $l_i=1/\sqrt{2}$ for all beam splitters, according to Eq. (\ref{scalar}) we have $\langle f_{i1}|f_{i2}\rangle=1$. Considering also $\theta_3=\theta_4$, the probabilities $P_j$ of photon detection at the corresponding detector and $A_i$ of absorption at the corresponding bean splitter are 
\begin{eqnarray}
	&&P_1=P_2=P_3=P_4=\frac{1}{16}\left[1+\cos(\phi+\theta_3)\right],\\
	&&A_a=A_b=\frac{1}{4}\,,\;A_c=A_d=\frac{1}{8}\left[1-\cos(\phi+\theta_3)\right],
\end{eqnarray}
which sum to 1 and present perfect absorption when $\phi+\theta_3$ is an odd multiple of $\pi$. The above $P_j$ agree with the results presented in Ref. \cite{jeffers19}, while the above $A_i$ are not computed in this reference. Note that the probability of absorption by the pair of beam splitters $a$ and $b$ is a constant, independent of the initial photon state and of the interferometer phases, such that there is no coherent nonlocal absorption by these objects as argued in Ref. \cite{jeffers19}. What happens is a local coherent absorption by the beam splitters $c$ and $d$, depending on the relative phases of the photon modes that arrive at their input ports, as seen in Eq. (\ref{psi2}).  The same happens in the case of the initial two-photon NOON state considered in Ref. \cite{jeffers19}.

This work was supported by the Brazilian agencies CNPq, CAPES, FAPEMIG, and CNPq/INCT-IQ (465469/2014-0).

\end{document}